\documentclass{sig-alternate}
\usepackage[utf8]{inputenc}

\begin{document}
% \setcopyright{acmcopyright}
\conferenceinfo{To appear at HealthRecSys'16,}{September 15, 2016}
\title{Extracting Food Substitutes From Food Diary via Distributional Similarity}
\numberofauthors{2}
\author{
\alignauthor
Palakorn Achananuparp\\
    \affaddr{Singapore Management University}\\
    \affaddr{Singapore}\\
    \email{palakorna@smu.edu.sg}
\alignauthor
Ingmar Weber\\
    \affaddr{Qatar Computing Research Institute}\\
    \affaddr{Qatar}\\
    \email{iweber@qf.org.qa}
}
\maketitle

\begin{abstract}
In this paper, we explore the problem of identifying substitute relationship between food pairs from real-world food consumption data as the first step towards the healthier food recommendation. Our method is inspired by the distributional hypothesis in linguistics. Specifically, we assume that foods that are consumed in similar contexts are more likely to be similar dietarily. For example, a turkey sandwich can be considered a suitable substitute for a chicken sandwich if both tend to be consumed with french fries and salad. To evaluate our method, we constructed a real-world food consumption dataset from MyFitnessPal's public food diary entries and obtained ground-truth human judgements of food substitutes from a crowdsourcing service. The experiment results suggest the effectiveness of the method in identifying suitable substitutes.

\end{abstract}





% Introduction
\section{Introduction}\label{sec:introduction}
Forming and maintaining healthy eating habits is important to individuals' long-term physical well-being. However, despite the availability of numerous dietary guidelines, few people are able to do so as demonstrated by the prevalence of chronic diseases such as obesity and type-2 diabetes. Arguably, part of the reason for such failure is that these guidelines are one-size-fits-all suggestions, making them difficult to be adopted habitually by individuals. In contrast to dietary guidelines, suggestions tailored to specific individuals from recommender systems may be more effective at facilitating incremental behavior change. More specifically, by learning about users' dietary behavior through data from mobile food consumption tracking apps such as MyFitnessPal (MFP), the systems can nudge the users towards ``similar but healthier'' alternatives by recommending food substitutes personalized to the users' current dietary needs and preferences. In this work, we explored a data-driven approach to extracting food substitutes from personal food consumption data as the first step into the healthier food recommendation. Thanks to the rise of self-monitoring practices enabled by mobile and wearable technology, we turned to a wealth of public food consumption data created by MFP users.

\begin{figure}[t]
\centering
\includegraphics[width=\columnwidth]{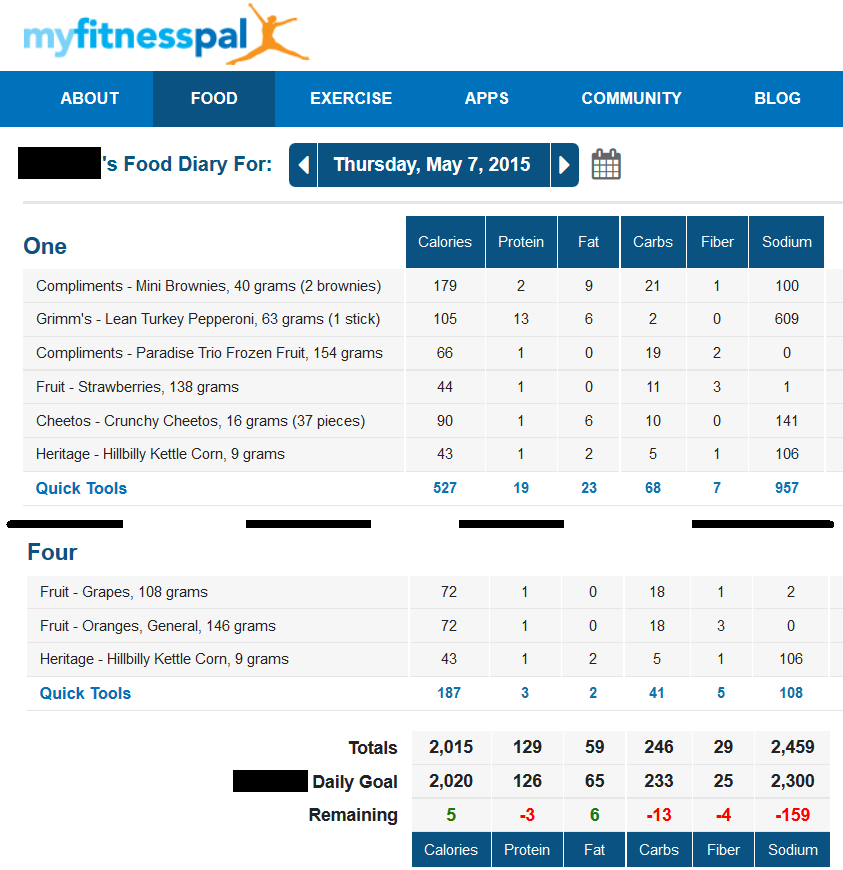}
\caption{Screenshot of a food diary on MFP.}
\label{fig:mfp_screenshot}
\end{figure}

MyFitnessPal (MFP) is a popular mobile app and website for fitness and health with 80 million registered users\footnote{\url{https://en.wikipedia.org/wiki/MyFitnessPal}}. One of its core features is an online food diary which helps users track their food consumption to achieve specific health goals, such as losing, gaining, or maintaining weight. Each food diary page consists of a sequence of meals where each meal contains a collection of food entries and nutrition information. As we can see from the sample diary page shown in Figure~\ref{fig:mfp_screenshot}, the user has logged 9 food entries across 2 meals (named ``One'' and ``Four''). When logging a new entry into a diary, users can either enter a new food entry and nutritional values or search for existing food entries shared by other users in the food database. Users can control the  diary sharing setting such that their diaries can be viewed by anyone (``Public''), their friends (``Friends Only''), or only the users themselves (``Private'').

Our main assumption of the substitution relationship between foods is inspired by the distributional hypothesis in linguistics: words that occur in the same contexts tend to convey similar meanings. By applying the same notion to food consumption, we hypothesize that foods consumed in similar contexts are more likely to be a substitute of each other. More specifically, our method is based on the vector space models of semantics~\cite{Turney2010} commonly used in related natural language processing (NLP) tasks, such as word similarity and analogy recovery.

Past studies have investigated the applications of recommender systems in food and cooking domains. One of the most common tasks explored by researchers is cooking recipe recommendation~\cite{Freyne:2010:IFP:1719970.1720021, Harvey2013, Ge2015} where popular recommendation algorithms, such as collaborative filtering~\cite{Freyne:2010:IFP:1719970.1720021} and matrix factorization~\cite{Harvey2013, Ge2015} were employed to predict ratings of cooking recipes. Others have focused on extracting substitutable ingredients from recipes using network analysis~\cite{Teng2012} and statistical approaches~\cite{ Boscarino2014}. While many past studies relied on recipe ratings data collected through recipe-sharing websites such as AllRecipes.com, to the best of our knowledge, our work is the first to study the food substitute extraction problem using a real-world self-reported food consumption data. In addition, we identified the substitution relationship directly from the food consumption data instead of relying on external knowledge sources\cite{Teng2012, Boscarino2014}. Lastly, we evaluated the effectiveness of the method on the human-labeled dataset of food substitutes constructed through an online crowdsourcing service.

The rest of the paper is organized as followed. First, we describe our method in Section~\ref{sec:method}. In Section~\ref{sec:data}, we describe the procedures to collect and process data. Then, we present the experimental evaluation in Section~\ref{sec:experiments} and discuss the results in Section~\ref{sec:results}. Lastly, we conclude the paper in Section~\ref{sec:conclusion}.

% Rule-based approach to extracting substitutable ingredients from recipes \cite{Shidochi2009}
% Van Pinxteren et al.\cite{VanPinxteren2011} proposed recipe similarity measure as simple euclidean distance between two recipe vectors
% Matrix factorization \cite{Forbes2011}
% Ahn et al.\cite{Ahn2011}
% Jain et al.\cite{Jain2015} analyzed food pairing
% Said and Bellogin\cite{Said2014} collected rating data from AllRecipes.com and analyze correlation between county-level obesity rate and ingredients
% Howell et al.\cite{Howell2016} predicted food tastes
% Rahman et al.\cite{Rahman2016} predicted eating moment

% Distributional Similarity
% Harris\cite{Harris1954} postulated the distributional hypothesis that words that occur in the same contexts tend to have similar meanings
% Church and Hank\cite{Church1990} introduced the use of PMI in distributional similarity models
% Turney\cite{Turney2001} introduced PMI-IR
% Bullinaria and Levy\cite{Bullinaria2007} showed that PPMI outperforms other weighting approaches
% Turney and Pantel\cite{Turney2010} reviewed VSMs for semantics, e.g. word-context matrix
% Malamud et al.\cite{Melamud2015} propose a simple model for lexical substitution, which is based on the popular skip-gram word embedding model

% On Statistical significance of PMI
% Damani and Ghonge\cite{Damani2013} 
% Washtell and Markert\cite{Washtell2009}

% Nutrient Rich score\cite{Drewnowski2015}

% The importance of substitutes and complements in recommender systems
% Zheng et al.\cite{Zheng2009}
% McAuley et al.\cite{McAuley2015}

% Methodology
\section{Food Substitute Extraction}\label{sec:method}
Our approach to food substitute extraction is based on the assumption that foods consumed in similar contexts tend to be similar dietarily, i.e., they are a suitable substitute of each other. In this work, the contexts comprise other foods consumed together in the same meals. Specifically, the substitutability between two foods is measured by the cosine similarity or dot product between their vector representations. We explore two explicit representation methods commonly used in similar NLP tasks~\cite{Levy2015}: Food-context matrix (PPMI matrix) and Singular Value Decomposition (SVD).

\subsection{Food-Context Matrix (PPMI Matrix)}

First, given the food consumption data, we constructed a food-context matrix $M$ where each row represents a food item $f \in V_f$ and each column represents a context $c \in V_c$, where $V_f$ and $V_c$ are the sets of observed food items and contexts, respectively. Each cell $M_{ij}$ represents the association between the food item $f_i$ and the context $c_j$ indicated by Positive Pointwise Mutual Information (PPMI) $ppmi_{ij}$ in Equation~\ref{eq:ppmi}. PPMI has been shown to perform better than other weighting approaches in semantic similarity tasks~\cite{Levy2015}.

% INGMAR: This is circular as pmi_{ij} is defined through a condition on pmi_{ij}
% \begin{equation}\label{eq:ppmi}
% pmi_{ij} =
% \begin{cases}
% \log{\frac{\#(f_i,c_j)*|D|}{\#(f_i)*\#(c_j)}} * \sqrt{\max({\#f_i,\#c_j})}, & \text{if } pmi_{ij} > 0\\
% 0 & \text{otherwise},
% \end{cases}
% \end{equation}

\begin{equation}\label{eq:ppmi}
ppmi_{ij} = \max(\log{\frac{\#(f_i,c_j)*|D|}{\#(f_i)*\#(c_j)}} * \sqrt{\max({\#f_i,\#c_j})}, 0)
\end{equation}

where $D$ denotes the set of all food-context pairs, $\#(f_i,c_j)$ denotes the number of times the pair ($f_i$, $c_j$) occurred in $D$, and $\#(f_i)$ and $\#(c_j)$ are the number of times $f_i$ and $c_j$ occurred in $D$, respectively. To obtain $D$, we define the contexts of food item $f_i$ as the food items that are consumed in the same meal as $f_i$. As PMI is known to bias towards infrequent events, we mitigate this problem by adopting a variant of $PMI_{sig}$ proposed in \cite{Damani2013}.

Finally, we measure the similarity between two food items by computing the cosine similarity between the corresponding row vectors in the food-context matrix $M$. Food pairs with higher cosine similarity are more likely to be a suitable substitute of each other than those with lower cosine similarity.

\subsection{Singular Value Decomoposition (SVD)}
% INGMAR: There is no matrix M defined anywhere. Probably matrix pmi?
One drawback of PPMI matrix is the sparsity of the food-context matrix $M$ which affects the performance of similarity measurements. One common way to improve the similarity computations is to perform dimensionality reduction through truncated Singular Value Decomposition (SVD) as proposed in Latent Semantic Analysis (LSA)~\cite{Deerwester1990}. Basically, SVD decomposes the matrix $M$ into the product of three matrices $U \Sigma V^T$, where $U$ and $V$ are orthogonal and $\Sigma$ is a diagonal matrix of singular values. Let $\Sigma_k$ be the diagonal matrix formed by selecting only the top $k$ singular values and $U_k$ and $V_k$ be the matrices formed by selecting the corresponding columns from $U$ and $V$, so that $M_k = U_k \Sigma_k V_k^T$ is a rank-$k$ approximation of $M$. Then, the similarity between two food items is measured from the dot product between two corresponding row vectors of $M_k$.

\section{Data Collection \& Processing}\label{sec:data}
In this section, we describe the procedures to collect food consumption data from MFP food diary pages, preprocess food entries in the diaries, and construct the food-context matrix and its low-dimensional representation.

\subsection{Obtaining Food Diaries}\label{sec:data:diaries}
We collected food diary data by web scraping MFP public food diary pages. First, we identified approximately 100,000 seed users who are a member of at least one of the 10 most popular groups in MFP communities. For each user whose food diary pages were publicly viewable, we retrieved up to the last 180 days of their food diaries (until March 2015 when the data collection took place). In total, 587,187 food diary pages of 9,896 users were retrieved. On average, each user has logged 59.3 days of diaries (S.D.\ = 54.6, median = 42) or 652.9 food entries in total (S.D.\ = 774, median = 366). The average age of users in the dataset is 35.6 years old (S.D. = 10.17). The vast majority of users are female (82\%) who live in the United States. The gender distribution of our dataset is similar to that of a larger sample used in \cite{Howell2016}.

\subsection{Preprocessing Food Entries}\label{sec:data:preprocessing}
After retrieving the food diaries, we parsed the HTML content of the diary pages to extract individual meals and food entries. Since food entries are described in free text, different entries may refer to the same dish. For example, both ``toasted tuna sandwich with cheese'' and ``grilled tuna sandwich'' are a kind of tuna sandwich. Furthermore, food entries often contain brand/restaurant name and serving size, e.g., ``Chili's - Santa Fe Chicken Salad, 3 cups''. Therefore, it is problematic to use the original text of the food entries in the analysis as two virtually identical entries might appear to be textually different.

To mitigate the problem, we represented each food entry as a set of salient features. A salient feature was extracted from the food entry text by matching word tokens of the food entry text with food-specific concepts, such as ingredients, preparation methods, etc. For this task, we manually built a food taxonomy\cite{Weber2016} consisting of main categories, subcategories, and entities (leaf nodes). For each main category in the taxonomy, we find the maximal match between taxonomic entities in that category and the food entry text. After a match was found, we created a salient feature by concatenating the corresponding main category, subcategory, and entity (in ``main category:subcategory:entity'' format) and added the term to the set of salient features. We found that the procedure was effective enough in removing most noises from the original food entry text. For example, after preprocessing, the food entry ``McDonald's - premium sweet chili chicken Wrap (grilled), 1 burger (200g)'' will be represented as a set of 3 salient features \{staple foods:wheat:wrap, meats:poultry:chicken, preparation methods:dry heat:grill\}. As a result, the number of unique food entries in the dataset was significantly reduced from 1.2 million to 71.7 thousand entries. About 10\% of food entries could not be matched to any entities and were subsequently discarded. Table~\ref{tab:dataset} summarizes the dataset\footnote{\url{https://goo.gl/Hkyi5w}} after the preprocessing steps.

\begin{table}
\centering
\begin{tabular}{ccc}
\hline
\# users & \# meals & \# unique food entries \\ \hline
9,896  & 1,919,024  & 71,7175 \\ \hline
\end{tabular}
\caption{MFP Dataset}
\label{tab:dataset}
\end{table}

\subsection{Building the Food-Context Matrix}\label{sec:data:matrix}
Finally, given the processed food diary data, we computed $PMI_{sig}$~\cite{Damani2013} for each pair of food entry $f \in V_f$ and context $c \in V_c$, where $c$ is another food entry occurred in the same meal as $f$, and built the food-context matrix $M$ with 22,804 rows ($|V_f|$) and 63,653 columns ($|V_c|$). Then, we applied LSA to the matrix $M$ to get the low-dimensional representation $M_k$. Typically, the value of $k$ is in the [500, 5000] range. In this work, we set $k=500$. 
% INGMAR: Can we say anything about why k=500 was chosen?

% The taxonomy has a reasonable coverage with respect to our dataset of food diaries. That is, out of 632,652 unique food entries, 88\% were successfully annotated with at least one category. The causes of failed annotation include users' input errors, such as misspellings of food names (e.g., brocolli, avacado, or spinich), entering food names that are too short or non-descriptive (e.g., ``Nestle - Fitness''), or non-English food names (e.g.\ huevos rancheros [Mexican dish], or char kway teow [Singaporean dish]).
 
%  To achieve this, we built a food taxonomy by compiling lists of food-related categories and page names from Wikipedia\footnote{http://en.wikipedia.org/wiki/Category:Foods}. The taxonomy is manually organized into 18 main categories (e.g., staple food, meats, vegetables, fruits, or preparation methods), 149 subcategories (e.g., wheat, rice, beef, chicken, or salad) and 4,233 entities (child nodes). Specifically, our goal is to use the taxonomy as a knowledge source to help automatically annotate each entry in a food diary with categories describing its ingredients and meal types. For example, the entry ``McDonald's - Premium Sweet Chili Chicken Wrap (Grilled)'' will be annotated with the following set of \{main category: subcategory: entity\}: \{Staple foods: Wheat: Wrap\}, \{Meats: Poultry: Chicken\}, \{Preparation Methods: Grill\}, \{Fast foods: McDonald's\}

% Experiments
\section{Experiments}\label{sec:experiments}
\subsection{Constructing the Evaluation Dataset}
We designed the evaluation of the food substitution extraction methods: PPMI matrix and SVD, as a top-$k$ substitute ranking task. To obtain human judgements of food substitutes, we used CrowdFlower\footnote{\url{http://www.crowdflower.com/}}, an online crowdsourcing service. First, we randomly chose 100 food entries containing ingredients from major protein groups (i.e., meats, beans and legumes, and nuts and seeds) to be used as target queries. Next, we generated a ranked list of top-10 food substitute candidates for each target query using each method. This resulted in 2,000 food substitute pairs (1,000 for each method) to be labeled by CrowdFlower workers. Then, we instructed each worker to rate how likely they agree that each food pair is a suitable substitute of each other on 7-point Likert scale responses from 1 (strongly disagree) to 7 (strongly agree). Each food pair was judged by 3 workers. For quality control, 57 test questions created by the first author were used as ground truths to filter out low-quality workers. Cohen's Kappa between the workers' labels and the ground truth labels was 0.87, indicating strong agreement.

\subsection{Evaluation Metrics}
We employed 3 metrics used in standard evaluation of ranked lists in information retrieval: precision at $k$ where $k=1$ and $k=10$, mean average precision (MAP), and normalized discounted cumulative gain (NDCG). To obtain the ground truth judgement for each food pair, we simply took the average of all ratings given by workers to the food pair. Since prec@k and MAP require binary judgements, we experimented with two binary judgement threshold $\tau$ values. Particularly, we were interested in comparing the performances when average ratings were greater than 3 (i.e., at least 'not disagree') or greater than 4 (i.e., at least 'slightly agree'). Any food pairs whose average ratings satisfying the binary judgement threshold are considered a true substitute pair. Lastly, all metrics have values in the [0, 1] range.

% Results
\section{Results \& Discussion}\label{sec:results}
Table~\ref{tab:perf-3} shows the performance of PPMI matrix and SVD on the food substitute ranking task given $\tau=3$. Overall, the results show that the vector space models can be effectively applied to extract food substitutes from food diaries. Both methods are equally good at identifying food substitutes according to prec@1, prec@10, and MAP. Next, with a more stringent threshold ($\tau=4$), SVD greatly outperforms PPMI matrix according to prec@1 (+18.97\%), prec@10 (+22.75\%), and MAP (+12.22\%). This is not surprising as SVD has shown to improve the similarity measurements in similar NLP tasks~\cite{Levy2015}. Interestingly, PPMI matrix is slightly better than SVD at generating ideal ranked lists of food substitutes according to NDCG (+5\%). Examples of top-10 substitutes for the food entry ``Tim Bacon - Bacon, 1 slices (54g)'' extracted by PPMI matrix are shown in Table \ref{tab:examples}. Evidently, the algorithm ranked food entries containing processed meats (e.g., sausages and bacon) as suitable substitutes for bacon slices higher than other protein groups.

\begin{table}[tbp]
\centering
\begin{tabular}{lllll}
\hline
Method & prec@1 & prec@10 & MAP & NDCG   \\ \hline
%PPMI Matrix & 0.75 & 0.777   & 0.8258 & 0.8105 \\
%SVD  & 0.77  & 0.777   & 0.8226 & 0.7719 \\ \hline
PPMI Matrix & 0.75 & 0.777   & 0.826 & 0.811 \\
SVD  & 0.77  & 0.777   & 0.823 & 0.772 \\ \hline
\end{tabular}
\caption{Performance of the each method ($\tau=3$).}
\label{tab:perf-3}
\end{table}

\begin{table}[tbp]
\centering
\begin{tabular}{llll}
\hline
Method & prec@1 & prec@10 & MAP\\ \hline
PPMI Matrix & 0.58 & 0.567  & 0.673 \\
SVD  & 0.69  & 0.696   & 0.755 \\ \hline
\end{tabular}
\caption{Performance of the each method ($\tau=4$). Since NDCGs are not affected by the threshold, they are omitted from the table.}
\label{tab:perf-4}
\end{table}

\begin{table}[tp]
\centering
\begin{tabular}{l}
\hline
Food entry\\ \hline 
Homemade - Sausage Balls, 8 -inch Ball\\
Sainsburys - Smoked Streaky Bacon Rashers, 2 rasher\\
Kroger - Traditional Cut Bacon, 2 slices\\
Pork Sausage, Spicy - Natures Promise, 1 Link\\
Pork - Cured, bacon, cooked, pan-fried, 2 slice cooked\\
Leidy's - Maple Glazed Premium Sliced Bacon, 1 Strips\\
Oscar Mayer - Turkey Bacon, 1 slice\\
Unknown - 2 Rasher of Grilled Bacon, 70 g\\
Bacon - Bacon Slices-oven Baked, 4 oven baked\\
Hormel - Black Label Bacon Original, 2 Pan Fried Slices\\ \hline
\end{tabular}
\caption{Top-10 food entries identified as substitutes for ``Tim Bacon - Bacon, 1 slices (54g)'' extracted by PPMI matrix.}
\label{tab:examples}
\end{table}

% Our study also shed light on the dietary behavior of the MFP users. Figure~\ref{fig:sub_heatmap} summarizes the co-occurrences of subcategories from 19,040 food substitute pairs containing ingredients from the major protein groups; the darker the cell color, the greater the co-occurrence. As we can see, the most frequently consumed substitutes were foods in the poultry subcategory such as chicken and turkey. Next, foods containing nuts were mostly substituted with other nuts. Lastly, apart from nuts, other plant-based proteins (e.g., pseudocereals, legumes, etc.) were rarely consumed in place of meat-based proteins.

\begin{figure}[tbp]
\centering
\includegraphics[width=\columnwidth]{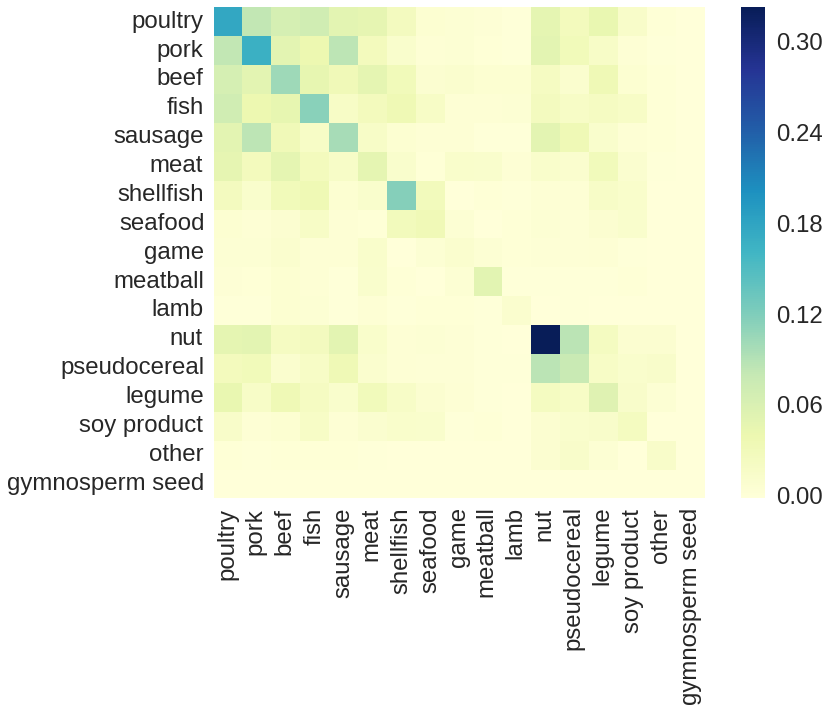}
% \caption{Co-occurrences of protein subcategories.}
\caption{Normalized co-occurrences of substitutes.}
\label{fig:sub_heatmap}
\end{figure}

Our study also shed light on the overall dietary behavior of the MFP users. Figure~\ref{fig:sub_heatmap} summarizes the normalized co-occurrences of food subcategories from 19,040 substitute pairs containing ingredients from the major protein groups. Each cell represents the Jaccard normalization of co-occurrence of two food subcategories; the darker the cell color, the greater the co-occurrence. As we can see, most substitutions were between foods in the poultry subcategory, such as chicken and turkey, and other subcategories. Next, foods containing nuts were mostly substituted with nuts and other plant-based proteins. Lastly, plant-based proteins were rarely consumed in place of meat-based proteins.

% Conclusion
\section{Conclusion}\label{sec:conclusion}
This study investigated the task of extracting food substitutes from the self-reported food consumption data. Our approach is based on the assumption that foods consumed in similar contexts are more likely to be a suitable substitute of each other. We applied the vector space models of semantics, commonly used in NLP, to identify food substitutes and created ground truth judgements of 2,000 food substitute pairs to evaluate the effectiveness of the methods. The experiments showed promising results. Our work is not without some limitations. First, because the majority of our CrowdFlower workers were from countries outside the US (i.e., Indonesia, Venezuela, and India), their judgements could be affected by cultural biases when labeling food consumption data created by users in the US. Next, our data preprocessing steps can be further improved. As previously discussed, about 10\% of all food entries had to be discarded due to the lack of salient features. For future work, we plan to experiment with other dense representation methods such as neural embeddings, incorporate higher-order co-occurrence and other contextual information, and identify ``personalized substitutes''. Lastly, to suggest ``similar but healthier'' options, we would also consider quantifying the healthfulness of foods through nutrient profiling.

% Acknowledgements
\section{Acknowledgements}\label{sec:ack}

This work is supported by the National Research Foundation under its International Research Centre @ Singapore Funding Initiative and administered by the IDM Programme Office.

% References
\bibliographystyle{abbrv}
\bibliography{main}

\end{document}